\documentclass[showpacs,preprintnumbers,superscriptaddress]{revtex4}
\usepackage{amsmath,amssymb,graphicx,bm}

\begin{document}

\title{The evolution of the power law k-essence cosmology}
\author{Rongjia Yang\footnote{Corresponding author}}
\email{yangrongjia@tsinghua.org.cn}
\affiliation{College of Physical Science and Technology, Hebei University, Baoding
071002, China}
\affiliation{Department of Physics, Tsinghua University, Beijing 100084, China}
\author{Bohai Chen}
\affiliation{College of Physical Science and Technology, Hebei University, Baoding
071002, China}
\author{Jun Li}
\affiliation{College of Physical Science and Technology, Hebei University, Baoding
071002, China}
\author{Jingzhao Qi}
\affiliation{Department of Physics, Institute of Theoretical Physics, Beijing Normal University, Beijing, 100875, China}

\begin{abstract}
We investigate the evolution of the power law k-essence field in FRWL
spacetime. The autonomous dynamical system and critical points are obtained.
The corresponding cosmological parameters, such as $\Omega _{\phi }$ and $w_{\phi }$, are calculated at these critical points.
We find it is possible to achieve an equation of state crossing through $-1$ for k-essence field.
The results we obtained indicate that the power law k-essence dark energy model can be compatible with observations.

\end{abstract}

\maketitle

\section{Introduction}
Since the cosmological constant model ($\Lambda$CDM) suffers from
cosmological constant problem \citep{Carroll:2000fy} as well as age problem \citep{Yang:2009ae}, many dynamic dark energy models have been proposed over the past
years, such as quintessence, k-essence, phantom, tachyon, etc. These scalar
field models can be seen as special cases of a model with Lagrangian, $\mathcal{L}_{\phi}=V(\phi)F(X)-f(\phi)$, with the kinetic energy
$X\equiv -\frac{1}{2}\partial_{\mu}\phi\partial^{\mu}\phi$ \citep{Carroll:2003st, Malquarti:2003nn}. This general Lagrangian has attracted much
attention. For some special cases, constraints have been
considered in \citep{Yang:2009zzl, Yang:2008hda, Yang:2008zzd, Yang:2010st} and
dynamics have been analyzed in \citep{Yang:2010vv, Yang:2012tm, Yang:2012ht, DeSantiago:2012nk}.
Geometrical diagnostic methods have been used to discriminate a class of this
theory from $\Lambda$CDM \citep{Gao:2010ia}. Unified model of inflation,
dark matter and dark energy have been discussed in \citep{Bose:2009kc, DeSantiago:2012nk, Saitou:2011hv}. Generalized tachyon models have been investigated
in \citep{Unnikrishnan:2008ki, Yang:2012ht}. Here we consider a model with $F(X)=-\sqrt{X}+X$, $V(\phi)\propto 1/\phi^2$, and $f(\phi)=0$. This type of k-essence has been shown to be a phenomenologically acceptable and theoretically interesting model which can unify inflation, dark matter, and dark energy \citep{Bose:2009kc, DeSantiago:2011qb}. We investigate the
possible cosmological behavior of this model in Friedmann-Robertson-Walker-Lema\^{i}tre (FRWL) spacetime by performing a phase-space and stability analysis. We
calculate various observable quantities, such as the density of the dark
energy and the equation of state (EoS) parameter in these solutions. The results show that the
model discussed here can be consistent with observations.

This paper is organized as follows: in the following section, we review
k-essence dark energy models. In the third section, we consider the dynamics
of the k-essence scaler field. In the fourth section, we discuss the stabilities of critical points and the model. Finally, we close with a few concluding
remarks in the fifth section.

\section{K-essence cosmology}
We consider k-essence dark energy models with Lagrangian
\begin{eqnarray}
\label{Lagrangians}
L=p_\phi=F(X)V(\phi),
\end{eqnarray}
where $F(X)$ and $V(\phi)$ are analytic functions of $X$ and $\phi$
respectively. $V(\phi)$ has dimension $M^4$ and hence $F(X)$ is dimensionless. Theses Lagrangians are invariant under the shift symmetry: $\phi\mapsto \phi+\phi_0$. Throughout this paper we will work with a flat, homogeneous, and isotropic
FRWL spacetime having signature $(-,+,+,+)$ and in units $c=8\pi G=1$. With Lagrangians (\ref{Lagrangians}), we can define the energy density, the pressure, and the EoS parameter. However, only after specifying the functional form of $F(X)$, it is possible to relate $F(X)$ with scale factor and then other dynamic quantities such as the energy density and the EoS parameter (recently, an interesting method, the effective field theory, for the most general class of single scalar field dark energy models has been developed in \citep{Gubitosi:2012hu,Bloomfield:2012ff,Gleyzes:2013ooa,Bloomfield:2013efa,Piazza:2013pua} in which perturbations are determined uniquely by only three time-dependent functions). Here we are interested in the power law k-essence with $F(X)=-\sqrt{X}+X$ which has been
extensively studied in \citep{Bose:2009kc, DeSantiago:2011qb} and has been shown to be phenomenologically acceptable and theoretically interesting model which can unify inflation, dark matter, and dark energy. The corresponding energy
density $\rho_\phi$, the EoS parameter $w_\phi$, and the effective sound
speed $c^{2}_{\mathrm{s}}$ are respectively given by
\begin{eqnarray}  \label{c}
\rho_\phi &=&V(\phi)[2XF_{X}-F]=XV, \\
w_\phi&=&\frac{F}{2XF_{X}-F}=\frac{X-\sqrt{X}}{X}, \\
c^{2}_{\mathrm{s}}&=&\frac{\partial p/\partial X}{\partial\rho /\partial X}=
\frac{F_{X}}{F_{X}+2XF_{XX}}=1-\frac{1}{2\sqrt{X}},
\end{eqnarray}
where $F_{X}\equiv dF/dX$ and $F_{XX}\equiv d^{2}F/dX^{2}$. The sound speed comes from the equation describing the evolution of
linear adiabatic perturbations in a k-essence dominated universe \citep{Garriga:1999vw} (non-adiabatic perturbation of k-essence has been discussed in \citep{Unnikrishnan:2010ag, Christopherson:2008ry}, we only consider here the case of adiabatic perturbation).

Since we only care the later evolution of the universe, we neglect baryonic
matter and radiation in the matter component. Then the Friedmann equations
take the form
\begin{eqnarray}  \label{f1}
H^2=\frac{1}{3}(\rho_{\mathrm{m}}+\rho_{\phi}), \\
\label{f2}
\dot{H}=-\frac{1}{2}(\rho_{\mathrm{m}}+\rho_{\phi}+p_{\phi}).
\end{eqnarray}
where $H=\dot{a}/a$ is the Hubble parameter. The equation of motion for the
k-essence field is given by
\begin{eqnarray}
\label{kf}
(F_{X}+2XF_{XX})\ddot{\phi}+3HF_{X}\dot{\phi}+(2XF_{X}-F)\frac{V_{\phi}}{V}%
=0,
\end{eqnarray}
where $V_{\phi}\equiv dV/d\phi$. Eqs. (\ref{f1}) and (\ref{f2}) are usually
transformed into an autonomous dynamical system when performing the
phase-space and stability analysis

\section{The basic equations and the critical points}
We introduce auxiliary variables
\begin{eqnarray}
x=\dot{\phi},~~~~y=\frac{\sqrt{V(\phi)}}{\sqrt{3}H},
\end{eqnarray}
to transform the cosmological equations (\ref{f1}) and (\ref{f2}) into an autonomous dynamical system as following
\begin{eqnarray}  \label{a1}
x^{\prime}&=&\frac{\sqrt{3}}{2}\lambda x^{2}y-3x+\frac{3\sqrt{2}}{2}, \\
\label{a2}
y^{\prime}&=&\frac{1}{4}y\left(-2\sqrt{3}\lambda xy+6-3\sqrt{2}
xy^{2}+3x^{2}y^{2}\right),
\end{eqnarray}
for $x>0$. And for $x<0$, Eqs. (\ref{f1}) and (\ref{f2}) turn into
\begin{eqnarray}
\label{a3}
x^{\prime } &=&\frac{\sqrt{3}}{2}\lambda x^{2}y-3x-\frac{3\sqrt{2}}{2},\\
\label{a4}
y^{\prime } &=&\frac{1}{4}y\left( -2\sqrt{3}\lambda xy+6+3\sqrt{2}
xy^{2}+3x^{2}y^{2}\right) ,
\end{eqnarray}
where the prime denotes a derivative with respect to the logarithm of the
scale factor, $\ln a$, and $\lambda \equiv -V_\phi/V^{\frac{3}{2}}$. Here we
are interested in the case where $\lambda$ is a constant, meaning $%
V(\phi)\propto \phi^{-2}$. The density parameters of k-essence, the EoS, the
sound speed, and the total EoS are reformulated as, respectively,
\begin{eqnarray}  \label{sound}
\Omega_{\phi}&=&\frac{1}{2}x^2y^2, \\
w_\phi&=&1-\sqrt{2}\mid{x}\mid^{-1}, \\
c^{2}_{\mathrm{s}}&=&1-\frac{\sqrt{2}}{2}\mid{x}\mid^{-1}, \\
w_{\mathrm{t}}&=&\Omega_{\phi}w_\phi=\frac{1}{2}x^2y^2-\frac{\sqrt{2}}{2}%
y^2\mid{x}\mid.
\end{eqnarray}
Because $0\leq \Omega_{\phi}\leq1$, the auxiliary variable $x$ and $y$ are
constrained as $0\leq \frac{1}{2}x^2y^2\leq1$.

Equations (\ref{a1}) and (\ref{a2}), (\ref{a3}) and (\ref{a4}), form self-autonomous dynamical
systems which are valid in the whole phase-space, not only at the critical
points. The critical points $(x_{\mathrm{c}},y_{\mathrm{c}})$ of the autonomous system are obtained by
setting the left-hand sides of the equations to zero, namely by solving $\mathbf{X}^{\prime}=(x^{\prime}, y^{\prime})^T=0$. Eight critical points are
obtained in all, as shown in Tables \ref{crit1} and \ref{crit2} which we also present the necessary conditions for
their existence, as well as the corresponding cosmological parameters, $%
c^{2}_{\mathrm{s}}$, $\Omega_{\phi}$, $w_{\phi}$, and $w_{\mathrm{t}}$. With
these cosmological parameters, we can investigate the possible state of the
universe and discuss whether there exists an acceleration phase or not.

\section{Stability}
As shown in \citep{Yang:2010vv, Yang:2012ht}, the stability of the critical point does not mean the stability of the
model, so, we must investigate both the stability of the critical point and the stability of the model.

\subsection{Stability of critical points}
To discuss the stability of the critical point, we expand $\textbf{X}$=$\{x,y\}$
around the critical values $\textbf{X}_{\rm c}$=$\{x_c,y_c\}$ by setting $\{x,y\}^T=\{x_c,y_c\}^T+\mathbf{U}$ with the perturbational variables $\mathbf{U}$ (see, for example, Refs. \citep{Copeland:1997et, Yang:2010vv, Capozziello:2005tf, Leon:2009rc}). Up to the first order we
acquire $\mathbf{U}^{\prime}=\mathbf{M}\cdot \mathbf{U}$ with the matrix $\mathbf{M}$ determined by
\begin{equation}
\mathbf{M}=\left[
\begin{array}{ccc}
\frac{\partial x^{\prime }}{\partial x} &  & \frac{\partial x^{\prime }}{
\partial y} \\
\frac{\partial y^{\prime }}{\partial x} &  & \frac{\partial y^{\prime }}{
\partial y}%
\end{array}
\right] .  \label{15}
\end{equation}
The matrix $\mathbf{M}$ contains the coefficients of the perturbation equations, thus its eigenvalues determine the stability of the critical points.
For hyperbolic critical points, all the eigenvalues have real
parts different from zero: sink for negative real parts is stable, saddle
for real parts of different sign is unstable, and source for positive real parts is unstable. The conditions for the stability of the critical points are given by Tr $\mathbf{M}<0$ and $\det \mathbf{M}>0$.

\begin{figure}[tbp]
\includegraphics[width=10cm]{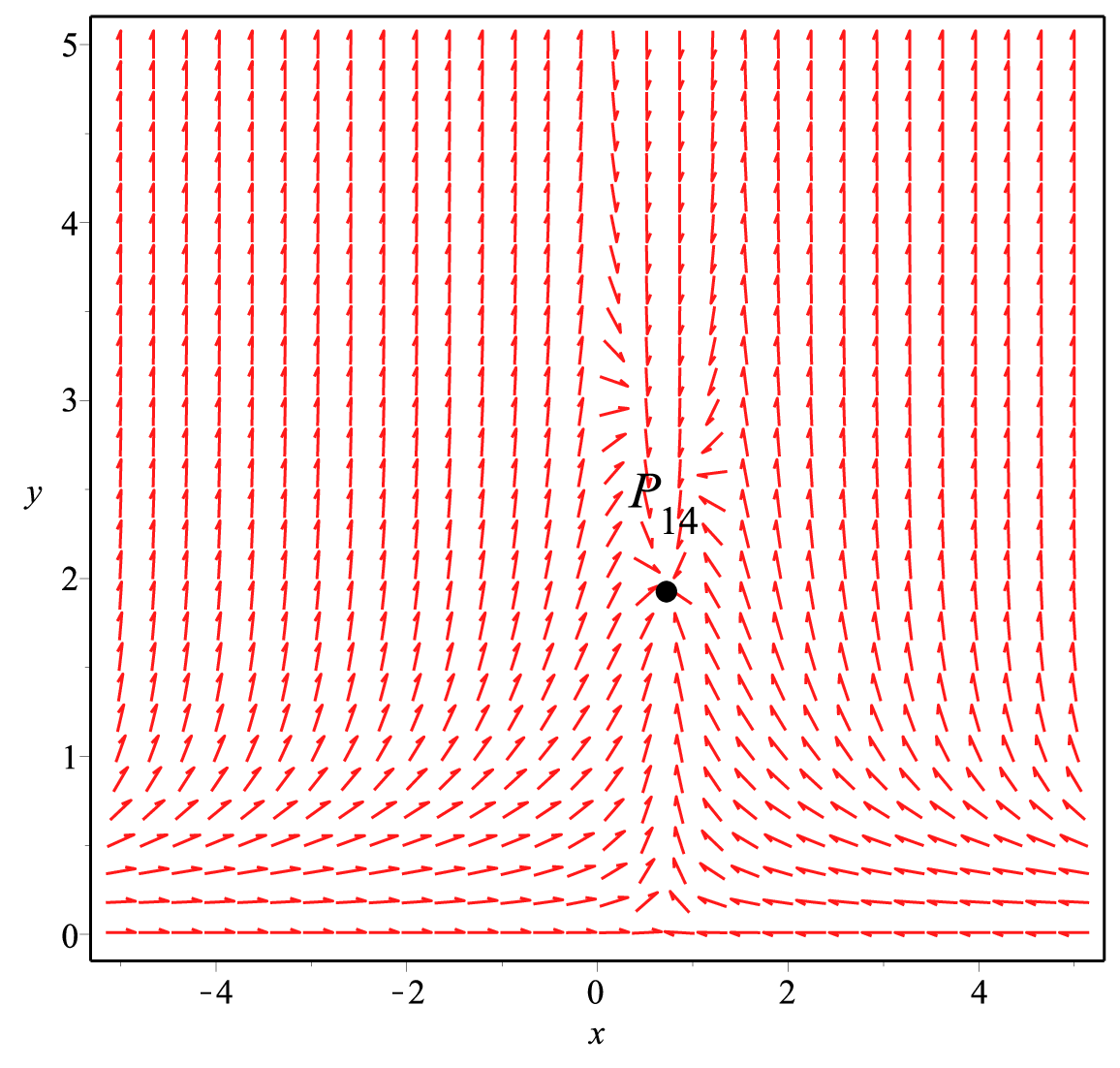}
\caption{Phase-space for power law k-essence cosmology, with the
choice $\protect\lambda =0.1$ for critical point $P_{14}$ when $x>0$. }
\label{Fig1}
\end{figure}

\begin{figure}[tbp]
\includegraphics[width=10cm]{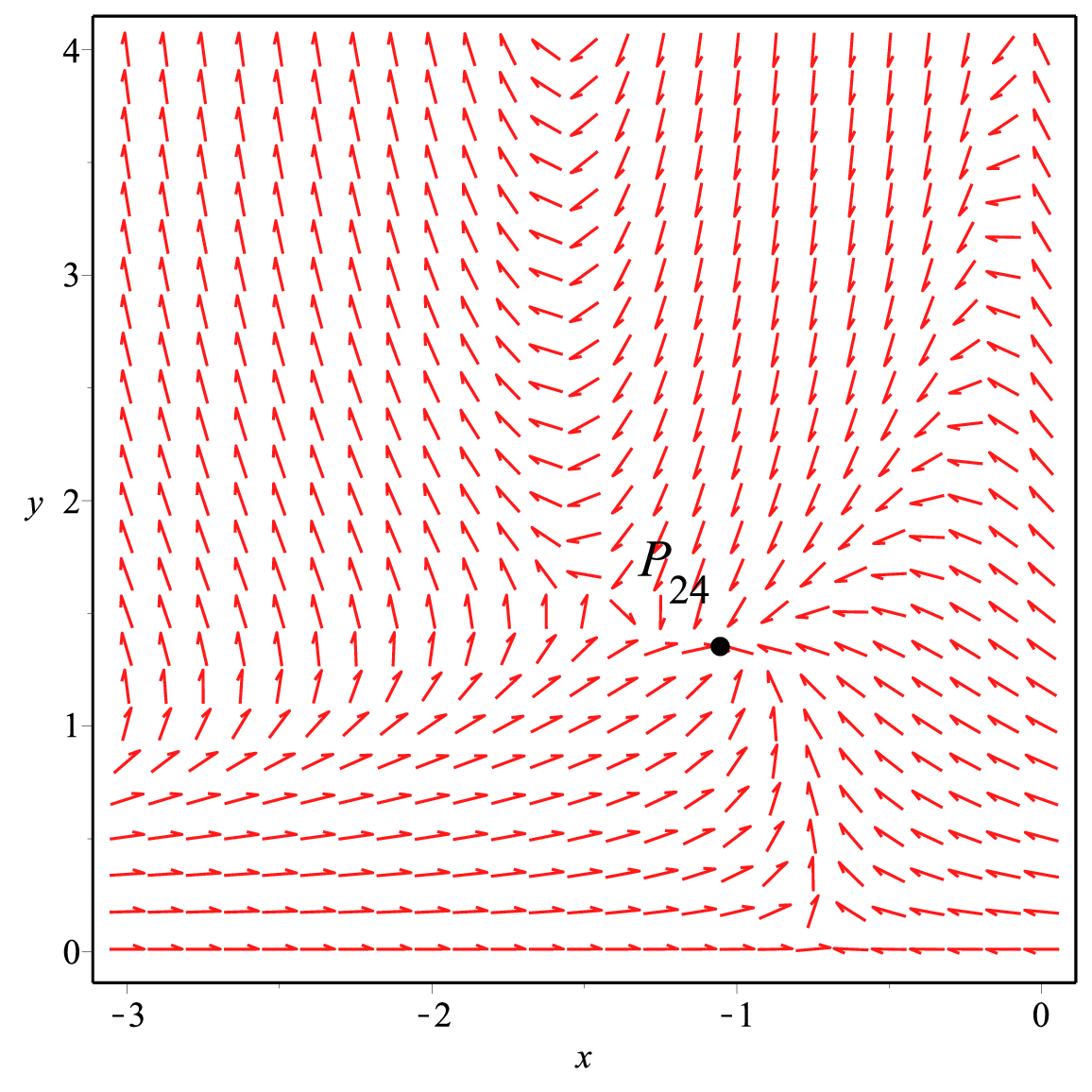}
\caption{Phase-space for power law k-essence cosmology, with the
choice $\protect\lambda =-0.8$ for critical point $P_{24}$ when $x<0$. }
\label{Fig4}
\end{figure}

For the power law k-essence dark energy we discussed here, the $\mathbf{M}$, $\det\mathbf{M}>0$, and Tr$\mathbf{M}<0$, are found to be
\begin{equation}
\mathbf{M}=\left[
\begin{array}{ccc}
\sqrt{3}\lambda xy-3 &  & \frac{\sqrt{3}}{2}\lambda x^{2} \\
\frac{1}{4}y(-2\sqrt{3}\lambda y-3\sqrt{2}y^{2}+6xy^{2}) &  & -\frac{\sqrt{3}%
}{2}\lambda xy+\frac{3}{2}-\frac{3\sqrt{2}}{4}xy^{2}+\frac{3}{4}x^{2}y^{2}+%
\frac{1}{4}y(-2\sqrt{3}\lambda x-6\sqrt{2}xy+6x^{2}y)%
\end{array}%
\right],
\end{equation}

\begin{eqnarray}
\det M &=&-\frac{9}{4}\lambda ^{2}x^{2}y^{2}+\frac{9\sqrt{3}}{2}\lambda xy-%
\frac{15\sqrt{6}}{8}\lambda x^{2}y^{3}+\frac{3\sqrt{3}}{2}\lambda x^{3}y^{3}-%
\frac{9}{2}+\frac{27\sqrt{2}}{4}xy^{2}-\frac{27}{4}x^{2}y^{2}, \\
\mathrm{~tr} M &=&\frac{\sqrt{3}}{2}\lambda xy-\frac{3}{2}-\frac{3\sqrt{2}}{4%
}xy^{2}+\frac{3}{4}x^{2}y^{2}+\frac{y}{4}(-2\sqrt{3}\lambda x-6\sqrt{2}%
xy+6x^{2}y),
\end{eqnarray}
for the case: $x>0$, and
\begin{equation}
\mathbf{M}=\left[
\begin{array}{ccc}
\sqrt{3}\lambda xy-3 &  & \frac{\sqrt{3}}{2}\lambda x^{2} \\
\frac{1}{4}y(-2\sqrt{3}\lambda y+3\sqrt{2}y^{2}+6xy^{2}) &  & -\frac{\sqrt{3}%
}{2}\lambda xy+\frac{3}{2}-\frac{3\sqrt{2}}{4}xy^{2}+\frac{3}{4}x^{2}y^{2}+%
\frac{1}{4}y(-2\sqrt{3}\lambda x+6\sqrt{2}xy+6x^{2}y)%
\end{array}%
\right],
\end{equation}
\begin{eqnarray}
\det M &=&-\frac{9}{4}\lambda ^{2}x^{2}y^{2}+\frac{9\sqrt{3}}{2}\lambda xy+%
\frac{15\sqrt{6}}{8}\lambda x^{2}y^{3}+\frac{3\sqrt{3}}{2}\lambda x^{3}y^{3}-%
\frac{9}{2}-\frac{27\sqrt{2}}{4}xy^{2}-\frac{27}{4}x^{2}y^{2}, \\
\mathrm{~tr} M &=&\frac{\sqrt{3}}{2}\lambda xy-\frac{3}{2}+\frac{3\sqrt{2}}{4%
}xy^{2}+\frac{3}{4}x^{2}y^{2}+\frac{y}{4}(-2\sqrt{3}\lambda x+6\sqrt{2}%
xy+6x^{2}y),
\end{eqnarray}
for the case: $x<0$.
\begin{table*}[tbp]
\caption{\label{crit1}For $x>0$, the existence and stability conditions of critical points, the cosmological parameters, and the range of $\lambda$ for acceleration.}
\label{crit1}
\begin{center}
\begin{tabular}{|l|l|l|l|l|l|l|l|}
\hline
Critical points $\{x_{c},y_{c}\}$ & Existence & stable & $c_{s}^{2}$ & $%
\Omega _{\phi }$ & $\omega _{\phi }$ & $\omega _{\rm tot}$ & Accelaration \\
\hline
$P_{11}=\{\frac{\sqrt{2}}{2},0\}$ & arbitrary & none & 0 & 0 & $-\frac{\sqrt{2}}{2}$ & 0
& none \\ \hline
$P_{12}=\{\sqrt{2},\frac{\sqrt{6}}{2\lambda }\}$ & $\lambda >0$ & $\lambda >
\frac{\sqrt{6}}{2} $ & $\frac{1}{2}$ & $\frac{3}{2\lambda ^{2}}$ & 0
& 0 & none \\ \hline
$P_{13}=\{\frac{\sqrt{3}}{\lambda+\sqrt{6}},-\frac{\sqrt{6}}{3}(
\lambda +\sqrt{6})\}$ & none & none & $-\frac{\sqrt{6}}{6}\lambda $
& 1 & $-1-\frac{\sqrt{6}}{3}\lambda $ & $-1-\frac{\sqrt{6}}{3}\lambda $ &
none \\ \hline
$P_{14}=\{\frac{\sqrt{3}}{\sqrt{6}-\lambda},\frac{\sqrt{6}}{3}(%
\sqrt{6}- \lambda)\}$ & $\lambda <\sqrt{6}$ & $\lambda<\frac{\sqrt{6}}{2}$ & $\frac{\sqrt{6%
}}{6}\lambda $ & 1 & $\frac{\sqrt{6}}{3}\lambda -1$ & $\frac{ \sqrt{6}}{3}%
\lambda -1$ & $ 0\leq\lambda<\frac{\sqrt{6}}{3}$ \\ \hline
\end{tabular}%
\end{center}
\end{table*}
\begin{table*}[tbp]
\caption{\label{crit2}For $x<0$, the existence and stability conditions of critical points, the cosmological parameters, and the range of $\lambda$ for acceleration.}
\label{crit2}
\begin{center}
\begin{tabular}{|l|l|l|l|l|l|l|l|}
\hline
Critical points $\{x_{c},y_{c}\}$ & Existence & stable & $c_{s}^{2}$ & $%
\Omega _{\phi }$ & $\omega _{\phi }$ & $\omega _{\rm tot}$ & Accelaration \\
\hline
$P_{21}=\{-\frac{\sqrt{2}}{2},0\}$ & arbitrary & none & 0 & 0 & -1 & 0 & none \\ \hline
$P_{22}=\{-\sqrt{2},-\frac{\sqrt{6}}{2\lambda }\}$ & $\lambda<0$ & $\lambda <-\frac{\sqrt{6}}{2} $ & $\frac{1}{2}$ & $\frac{3}{2\lambda ^{2}}$
& 0 & 0 & none \\ \hline
$P_{23}=\{\frac{\sqrt{3}}{\lambda-\sqrt{6}},\frac{\sqrt{6}}{3}(\lambda-\sqrt{6})\}$
& none & none & $\frac{\sqrt{6}}{6}\lambda $ & 1 & $-1+\frac{\sqrt{6}}{3}%
\lambda$ & $-1+\frac{\sqrt{6}}{3}\lambda$ & none \\ \hline
$P_{24}=\{-\frac{\sqrt{3}}{\lambda+\sqrt{6}},\frac{\sqrt{6}}{3}(\lambda+\sqrt{6})\}$
& $\lambda>-\sqrt{6}$ & $\lambda >-\frac{\sqrt{6}}{2}$ & $-%
\frac{\sqrt{6}}{6}\lambda $ & 1 & $-1-\frac{\sqrt{6}}3{}\lambda$ & $-1-\frac{%
\sqrt{6}}{3}\lambda$ & $ -\frac{\sqrt{6}}{3}<\lambda\leq0$ \\ \hline
\end{tabular}%
\end{center}
\end{table*}
According to the conditions for the stability of critical points, we
obtain the ranges of $\lambda$ in which the critical points are stable, as shown
in Tables \ref{crit1} and \ref{crit2}. We plot critical point $P_{14} $ for $\lambda=0.1$ in Fig \ref{Fig1} and $P_{24}$ for $\lambda=-0.8$ in Fig \ref{Fig4} to have a visual understanding of the behavior of the field near critical points.

\subsection{Stability of model}
The stability of model includes classical and quantum stability. We first discuss the
classical stability. In a flat universe, the equation for the canonical quantization variable $v$
describing the collective metric and scalar field perturbations takes the form \citep{Garriga:1999vw}
\begin{eqnarray}
v^{\prime\prime}_k+(c^{2}_{\mathrm{s}}k^2-\frac{\Phi^{\prime\prime}}{\Phi}%
)v_k=0,
\end{eqnarray}
where $\Phi=a(\rho_\phi+p_\phi)^{1/2}/(c_{\mathrm{s}}H)$ with $H$ the Hubble parameter. The
increment of instability is inversely proportional to the wave-length of the
perturbations, therefore the background model is violently unstable and do not has any physical significance for $c^{2}_{\mathrm{s}}<0$. Another potentially interesting requirement is $c_{\mathrm{s}}^2 \leq 1$, saying that the sound speed should not exceed the speed of light, otherwise the causality will be violated. Note, however, this is still an open problem (see e. g. \citep{Bruneton:2006gf, Kang:2007vs, Bonvin:2006vc, Ellis:2007ic, Babichev:2007dw, Gorini:2007ta}). Here we take the conditions for classical stability as: $1 \geq c_{\rm s}^{2}\geq 0$, namely
\begin{eqnarray}  \label{cq}
1\geq 1-\frac{\sqrt{2}}{2}\frac{1}{|x|}\geq 0,
\end{eqnarray}
for the case of power law k-essence we discussed here. From this equation, we obtain the range of $\lambda$ in which the model is classically stable: $|x|\geq\frac{\sqrt{2}}{2}$.

%

Now discussions for the quantum stability of the k-essence field are in order.
Expanding $p$ at second order in $\delta \phi$, the Hamiltonian fluctuations are found to be \citep{ArmendarizPicon:2005nz, Bamba:2012cp, Kahya:2006hc, Piazza:2004df}:
\begin{eqnarray}  \label{fluc}
\delta \mathcal{H}=p_X \frac{(\nabla \delta \phi)^2}{2}+(p_X+2Xp_{XX})\frac{(\delta \dot{\phi})^2}{2}-p_{\phi\phi}\frac{(\delta \phi)^2}{2},
\end{eqnarray}
where $p_{\phi\phi}\equiv d^2p/d\phi^2$. The positivity of the first two
terms in equation (\ref{fluc}) leads to the following conditions for quantum stability
\begin{eqnarray}
p_X \geq 0, ~~{\rm and} ~~p_X+2Xp_{XX} \geq 0.
\end{eqnarray}
The conditions for quantum stability for the power law k-essence dark energy discussed here are found to be: $|x|\geq\frac{\sqrt{2}}{2}$. Here $p_X \geq 0$ is the gradient-stability condition and $p_X+2Xp_{XX} \geq 0$ is the no-ghost condition. In general, violations of the null energy condition may lead to gradient instability. One way to avoid the gradient instability is to flip the sign of the kinetic term with a minimally coupled scalar field \citep{Caldwell:1999ew}, however, this turns
out to be catastrophic since the considered theory would inevitably develop ghost instabilities \citep{Cline:2003gs}, and as shown in Figures 1 and 2 in \citep{Piazza:2013pua}, a minimally coupled scalar cannot produce a super-accelerating equation of state. Another way is to consider the high order terms, and it has been shown that a quintessence model with $w_\phi\leqslant -1$ can be completely stable for some conditions \citep{Creminelli:2008wc,Creminelli:2006xe}. For k-essence or dark energy models crossing the phantom divide, the speed of sound should be set to zero to obtain stability \citep{Creminelli:2008wc}. For generalized and detailed discussions on the problem of the soundness of the theory against ghost-like and gradient instabilities we refer to reference \citep{Piazza:2013pua}. Discussions about perturbational instability for violating the null energy condition can also be found in \citep{Xia:2007km,Cai:2009zp,Cai:2012va,Dubovsky:2005xd,Rubakov:2014jja,Guo:2003rs}.

So it can be concluded that the model is both classically and quantum stable for $|x|\geq\frac{\sqrt{2}}{2}$. We say the model is (classically and quantum) stable at a critical point if its $x_{\mathrm{c}}$ is in the range of $x$ allowed by the conditions of stability for the model, or is not stable if $x_{\mathrm{c}}$ is not in the range of $x$ allowed by the conditions of stability for model \citep{Yang:2010vv, Yang:2012ht}.

\subsection{Cosmological implications}
For $x>0$, the model is stable at critical points $P_{13}$ for $-\sqrt{6}\leq \lambda \leq0$, $P_{14}$ for $0\leq \lambda \leq \sqrt{6}$, and $P_{11}$ and $P_{12}$ for arbitrary $\lambda$. But critical points $P_{11}$ is not stable and $P_{13}$ does not exist, so they are not relevant from a cosmological point of view. In other words, only critical points $P_{12}$ and $P_{14}$ are physical interesting.

For $\lambda>\frac{\sqrt{6}}{2}$, the critical point $P_{12}$ is stable. At this point, the k-essence behaves like dark matter with $\Omega_{\phi }=\frac{3}{2\lambda ^{2}}$, meaning the universe is partly occupied by k-essence. If $\lambda\longrightarrow +\infty$, the universe will be dominated by dark matter, while if $\lambda\longrightarrow \frac{\sqrt{6}}{2}$, the universe will be dominated by k-essence.

For $\lambda<\frac{\sqrt{6}}{2}$, the critical point $P_{14}$ is stable, while the range of $\lambda$ in which the model is stable is $0\leq\lambda\leq\sqrt{6}$, that is to say, only for $0\leq\lambda<\frac{\sqrt{6}}{2}$, both the model and the critical point are stable. At this point, the universe is dominated by k-essence with $\Omega_\phi=1$ and $w_\phi=\frac{\sqrt{6}}{3}\lambda-1$. If $\lambda=0$, the k-essence will behave like cosmological constant; while if $\lambda\longrightarrow \frac{\sqrt{6}}{2}$, the k-essence will behave like dark matter. The deceleration parameter is $q=-1+\frac{\sqrt{6}}{2}\lambda$. The final state of the universe dependents on the potential: the universe will speed up if $0\leq\lambda<\frac{\sqrt{6}}{3}$, will expand with constant-speed if $\lambda=\frac{\sqrt{6}}{3}$, and will speed down if $\frac{\sqrt{6}}{3}<\lambda< \frac{\sqrt{6}}{2}$.

\begin{figure}[tbp]
\includegraphics[width=10cm]{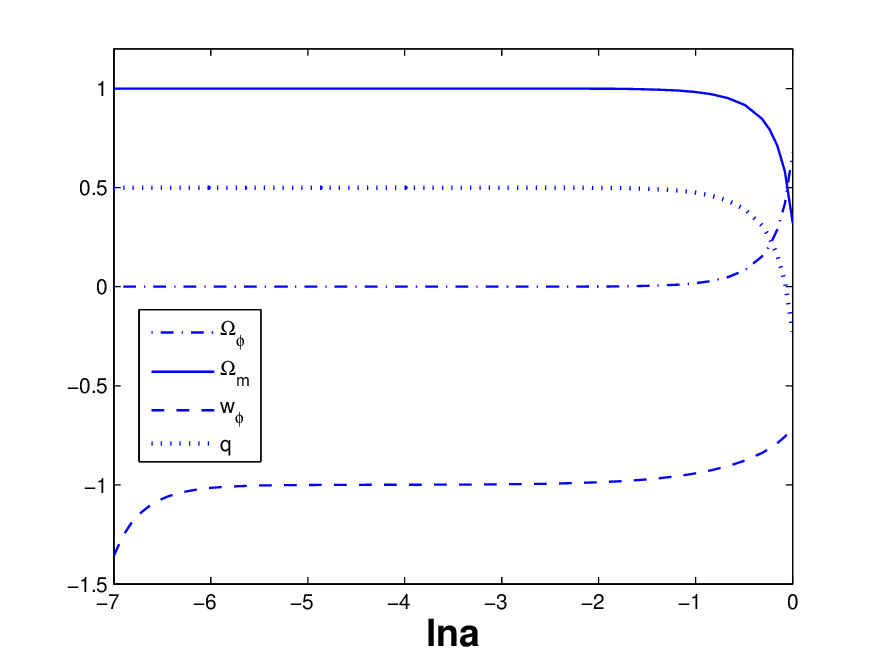}
\caption{The evolution of $\Omega_\phi$, $\Omega_{\rm m}$, $w_\phi$, and the deceleration parameter $q$ for $\lambda=0.1$ with the initial conditions $x=0.65$ and $y=0.0000375$ when $\ln a=-7$.}
\label{Fig5}
\end{figure}

We plot the evolution of $\Omega_\phi$, $\Omega_{\rm m}$, $w_\phi$, and the deceleration parameter $q$ for $\lambda=0.1$ (namely for the case $x>0$) in figure \ref{Fig5}. The initial conditions are chosen as $x=0.65$ and $y=0.0000375$ when $\ln a=-7$. In this case, an interesting result is that the EoS is smaller than $-1$ at early times and is larger than $-1$ at late times. The parameter $\Omega_\phi$ is nearly zero at early times and increase to $0.68$ when $\ln a\longrightarrow 0$, which is compatible with observations.

For $x<0$, the critical point $P_{21}$ is not stable and $P_{23}$ does not exist, while other critical points are stable for a certain range of $\lambda$. the model is stable at critical points $P_{23}$ for $\sqrt{6}\leq \lambda \leq 2\sqrt{6}$, $P_{24}$ for $-\sqrt{6}\leq \lambda \leq 0$, and $P_{21}$ and $P_{22}$ for arbitrary $\lambda$.

For $\lambda<-\frac{\sqrt{6}}{2}$, the critical point $P_{22}$ is stable, and the k-essence behaves like dark matter with $\Omega_{\phi }=\frac{3}{2\lambda ^{2}}$. If $\lambda\longrightarrow -\infty$, the universe will be dominated by dark matter, while if $\lambda\longrightarrow -\frac{\sqrt{6}}{2}$, the universe will be dominated by k-essence.

For $-\frac{\sqrt{6}}{2}<\lambda\leq0$, both the model and the critical point $P_{24}$ are stable. The universe is dominated by k-essence with $\Omega_\phi=1$ and $w_\phi=-\frac{\sqrt{6}}{3}\lambda-1$ at this point. If $\lambda=0$, the k-essence will behave like cosmological constant; while if $\lambda\longrightarrow -\frac{\sqrt{6}}{2}$, the k-essence will behave like dark matter. The deceleration parameter is $q=-1-\frac{\sqrt{6}}{2}\lambda$. The final state of the universe dependents on the potential: the expansion of universe will speed up if $-\frac{\sqrt{6}}{3}\leq\lambda<0$, will keep constant-speed if $\lambda=-\frac{\sqrt{6}}{3}$, and will speed down if $-\frac{\sqrt{6}}{2} <\lambda< -\frac{\sqrt{6}}{3}$.

\begin{figure}[tbp]
\includegraphics[width=10cm]{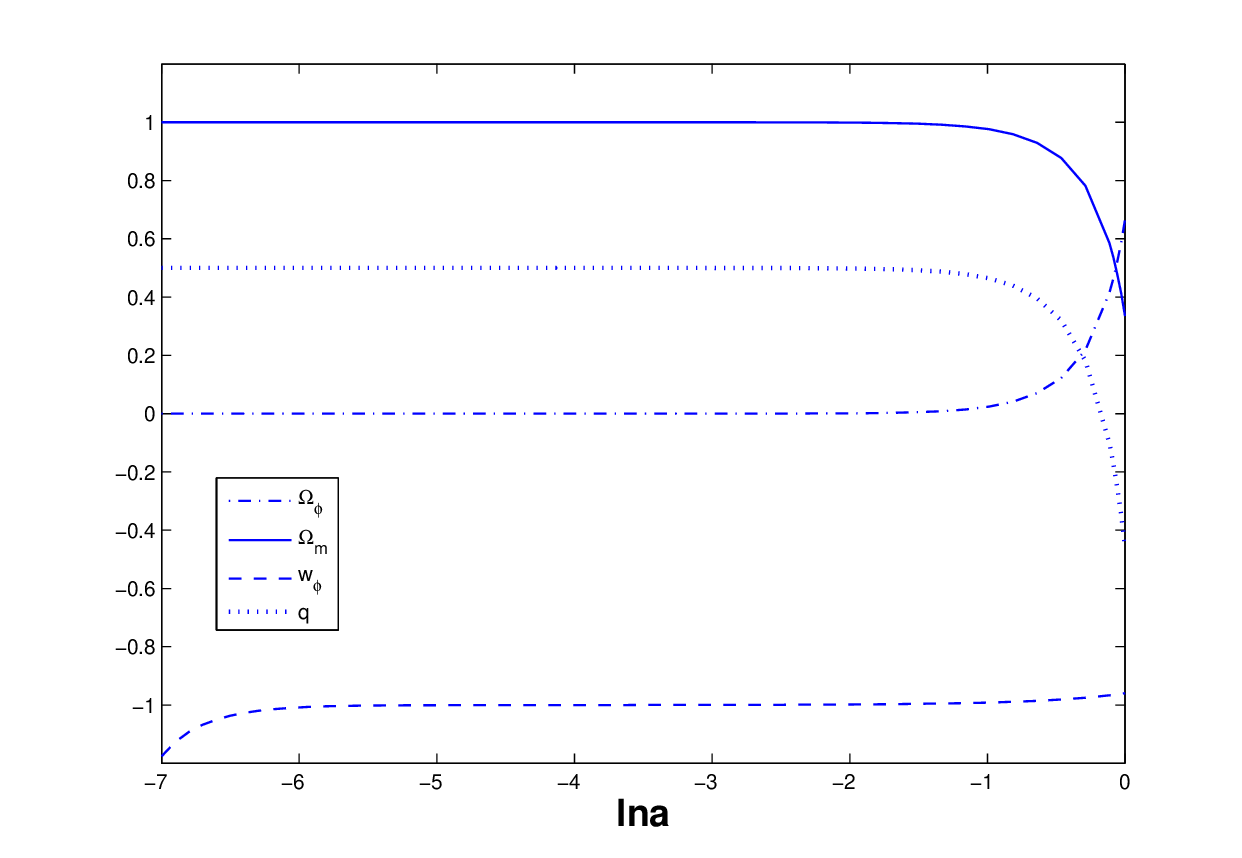}
\caption{The evolution of $\Omega_\phi$, $\Omega_{\rm m}$, $w_\phi$, and the deceleration parameter $q$ for $\lambda=-0.8$ with the initial conditions $x=-0.65$ and $y=0.000034$ when $\ln a=-7$.}
\label{Fig6}
\end{figure}

The evolution of $\Omega_\phi$, $\Omega_{\rm m}$, $w_\phi$, and the deceleration parameter $q$ for $\lambda=-0.8$ (namely for the case $x<0$) are plotted in figure \ref{Fig6} with the initial conditions $x=-0.65$ and $y=0.000034$ when $\ln a=-7$. The parameter $\Omega_\phi$ is nearly zero at early times and increase to $0.68$ when $\ln a\longrightarrow 0$, which is also compatible with observations. The EoS can also cross through $-1$ .

\section{Conclusions and discussions}
We have investigated the evolution of the universe when power law k-essence acts as dark energy and have examined whether there are late-time solutions compatible with observations.
Critical points and the conditions for their existence and stability are obtained. The corresponding cosmological parameters, $c_{\mathrm{s}}^{2}$, $\Omega _{\phi }$, $w_{\phi }$, and $w_{\mathrm{t}}$, are calculated at these critical points. The (classical and quantum) stability of the model are also discussed.

As discussed in \citep{Yang:2010vv, Yang:2012ht}, the stability of critical points does not mean the
stability of the model, vice versa. The critical points can be divided into three
classes: stable points at which the model is (classically
or quantum) unstable, stable points at which the model is stable, unstable points at which the model is stable \citep{Yang:2010vv, Yang:2012ht}.
From a cosmological point of view, only stable points at which the model is also
(classically and quantum) stable are physically interesting.

So for the case of $x>0$, only points $P_{12}$ and $P_{14}$ are cosmological
relevant. At the critical point $P_{14}$, the expansion of the universe can speed down, speed up,
or keep-constant speed. For the case of $x<0$, only points $P_{22}$ and $P_{24}$ are physically interesting. At the critical point $P_{24}$, the expansion of
the universe can also speed down, speed up, or keep constant-speed. The
final state of the universe dependents on k-essence field and its potential.
In both of these two cases, it is possible to have an EoS crossing through $-1$, this is an interesting result.

As it has been shown that in order to study the possible state of the power law k-essence cosmology, it is important to investigate both the stability
of the critical points and the (classical and quantum) stability of the
model \citep{Yang:2010vv, Yang:2012ht}. Otherwise the analysis will lead to wrong conclusions. The analysis we performed here indicates that the power law k-essence dark energy model can be compatible with observations. Theses results
can been taken into account if k-essence cosmology successfully passes observational tests which are interesting studies for other studies.

\begin{acknowledgments}
This study is supported in part by National Natural Science Foundation of China (Grant Nos. 11147028 and 11273010) and Hebei Provincial Natural Science Foundation of China (Grant No. A2011201147 and A2014201068).
\end{acknowledgments}

\bibliography{DK}
\end{document}